\documentclass{iopart}
\usepackage{amssymb}
\usepackage{iopams}
\usepackage[dvips]{graphicx}

\catcode`\@=11
\renewcommand\footnoterule{%
  \kern-3\p@
  \hrule\@width.4\columnwidth
  \kern2.6\p@}
\renewcommand\@makefntext[1]{%
    \parindent 1em\noindent
    \hb@xt@1.8em{\hss$^{\@thefnmark}$)}\hspace{2pt}%
    \footnotesize\rmfamily#1}  
\def\@makefnmark{\hspace{.5pt}\hbox{$^{\@thefnmark}$%
\hspace{-1pt})}} 

\def\RR{\mathbb{R}}

\def\CC{\mathbb{C}}

\def\PP{\mathbb{P}}

\def\ZZ{\mathbb{Z}}

\newcommand{\cC}{\mathcal{C}}

\newcommand{\cH}{\mathcal{H}}

\newcommand{\cM}{\mathcal{M}}
\newcommand{\cP}{\mathcal{P}}
\newcommand{\cT}{\mathcal{T}}

\newcommand{\cV}{\mathcal{V}}

\newcommand{\bfPhi}{\mathbf{\Phi}}
\newcommand{\bfPsi}{\mathbf{\Psi}}

\newcommand{\bfTheta}{\mathbf{\Theta}}

\newcommand{\bX}{\mathbf{X}}

\newcommand{\fS}{\mathfrak{S}}

\newcommand{\codim}{\mbox{\rm codim\,}}

\newcommand{\const}{\mbox{\rm const}\,}
\newcommand{\vspan}{\mbox{\rm span}\!}

\def\p{\partial}
\def\a{\alpha}
\def\b{\beta}
\def\d{\delta}
\def\g{\gamma}

\def\e{\varepsilon}

\def\lb{\lambda}

\renewcommand\Im{{\rm Im}}
\renewcommand\Re{{\rm Re}}

\newcommand{\be}[1]{\begin{equation}\label{#1}}
\newcommand{\ee}{\end{equation}}
\newcommand{\ba}[1]{\begin{eqnarray}\label{#1}}
\newcommand{\ea}{\end{eqnarray}}
\newcommand{\nn}{\nonumber}
\newcommand{\rf}[1]{(\ref{#1})}
\begin{document}

\title[Projective Hilbert space structures]{Projective Hilbert space structures at exceptional points}

\author{Uwe G\"unther$^a$, Ingrid Rotter$^b$ and Boris F. Samsonov$^c$\footnote{Deceased 08 November 2012}}

\address{$^a$ Helmholtz Center Dresden-Rossendorf,  Bautzner Landstra\ss e 400,
D-01328 Dresden, Germany}
\address{$^b$ Max Planck Institute for physics of complex systems,
D-01187 Dresden, Germany}
\address{$^c$ Physics Department, Tomsk State University, 36 Lenin Avenue,
634050 Tomsk, Russia}

\eads{ \mailto{u.guenther@hzdr.de} and \mailto{rotter@pks.mpg.de}}

\begin{abstract}
A non-Hermitian complex symmetric $2\times 2$ matrix toy model is
used to study projective Hilbert space structures in the vicinity
of exceptional points (EPs). The bi-orthogonal eigenvectors of a
diagonalizable matrix are Puiseux-expanded  in terms of the root
vectors at the EP. It is shown that the apparent contradiction
between the two incompatible normalization conditions with finite
and singular behavior in the EP-limit can be resolved by
projectively extending the original Hilbert space. The
complementary normalization conditions correspond then to two
different affine charts of this enlarged projective Hilbert space.
Geometric phase and phase jump behavior are analyzed and the
usefulness of the phase rigidity as measure for the distance to EP
configurations is demonstrated. Finally, EP-related aspects of
$\cP\cT-$symmetrically extended Quantum Mechanics are discussed
and a conjecture concerning the quantum brachistochrone problem is
formulated.
\end{abstract}
\pacs{03.65.Fd, 03.65.Vf, 03.65.Ca, 02.40.Xx}
{\small published as: \quad \ \ J. Phys. A: Math. Theor. {\bf 40}, 8815 (2007)}

\section{Introduction}

A generic property of non-Hermitian operators is the possible
occurrence of non-trivial Jordan-blocks in their spectral
decomposition \cite{kato}. For an operator $H(\bX)$ depending on a
set of parameters $\bX=(X_1,\ldots,X_m)\in\cM$ from a space $\cM$
this means that, in case of a single Jordan block, two or more
spectral branches $\lambda_1(\bX),\ldots, \lambda_k(\bX)$ may
coalesce (degenerate) at certain parameter  hypersurfaces
$\cV\subset\cM$ under simultaneous coalescence of the corresponding
eigenvectors $\Phi_1(\bX),\ldots,\Phi_k(\bX)$:
$\lambda_1(\bX_c)=\cdots=\lambda_k(\bX_c)$,
$\Phi_1(\bX_c)=\cdots=\Phi_k(\bX_c)\equiv\Theta_0(\bX_c)$ for
$\bX_c\in\cV$. Spectral points of this type are branch points of the
spectral Riemann surface and are called exceptional points (EPs)
\cite{kato}. At the EPs the set of the originally $k$ linearly
independent eigenvectors $\Phi_1(\bX),\ldots,\Phi_k(\bX)$ is
replaced by the single eigenvector $\Theta_0(\bX_c)$ and $k-1$
associated vectors $\Theta_1(\bX_c),\ldots,\Theta_{k-1}(\bX_c)$
which form a Jordan chain. Together they span the so called
$k-$dimensional algebraic eigenspace (or root space)
$\fS_\lb(\bX_c)=\vspan\,
[\Theta_0(\bX_c),\ldots,\Theta_{k-1}(\bX_c)]$ \cite{kato,baumg} so
that the total space dimension remains preserved. The construction
extends straight forwardly to the presence of several Jordan blocks
for the same eigenvalue $\lb(\bX_c)$. In general, the degeneration
hypersurface $\cV\subset\cM$ consists of components $\cV_a$ of
different codimension $\codim \cV_a=a$. Higher order Jordan blocks
require a higher degree of parameter tuning (they have a higher
codimension) and a correspondingly lower dimension of the component
$\cV_a$. Due to the different dimensions of its components $\cV_a$
the degeneration hypersurface $\cV=\bigcup\limits_a \cV_a$ itself
has the structure of a stratified manifold \cite{stratification}.

EPs occur naturally in quantum scattering setups
\cite{newton,mahaux-weid-1} when two or more resonance states
coalesce and higher order poles of the S-matrix form. Within the
Gamow state approach such S-matrix double poles have been considered
in
\cite{mondragon-lett0,mondragon-1,bohm-jordan,mondragon-s-matrix-2},
whereas in the Feshbach projection operator formalism (one of the
basic approaches to analyze open quantum systems)  they naturally
occurred in studies of nuclei \cite{rep}, atoms
\cite{marost23,vanroose} and quantum dots \cite{rs1,rs2}. EP-related
crossing and avoided crossing scenarios have been studied for bound
states in the continuum \cite{marost23,rs3,friedrich,brs1} as well
as for phase transitions
\cite{heiss-lipkin1,heiss-m-rotter,JMR,SSR}. In asymptotic analyses
of quasi-stationary systems EPs show up as hidden crossings
\cite{soloviev}. EP-related theoretically predicted level and width
bifurcation properties have been experimentally verified in a series
of microwave resonator cavity experiments. In \cite{prsb-prl2000},
the resonance trapping phenomenon (width bifurcation) \cite{rep} has
directly been proven.  The fourfold winding around an EP has been
found experimentally \cite{darmstadt} in full agreement with the
theoretical prediction \cite{heiss-m-rotter,heiss-01} and related
studies \cite{rs2,Korsch-Mossmann-1}. In
\cite{heiss-harney-chirality} two-level coalescences have been
associated with chiral system behavior. The geometric phase at EPs
has been discussed in
\cite{rs2,Korsch-Mossmann-1,heiss-harney-chirality,garrison,berry-naples,berry-non-herm1,mondrag-geom,MKS-1}.

EPs play also an important role in the recently considered
$\cP\cT-$symmetrically extended quantum models
\cite{CMB-PTSQM,CMB-brachist,CMB-rev}. There they correspond to
the phase transition points between physical sectors of exact
$\cP\cT-$symmetry and unphysical sectors of spontaneously broken
$\cP\cT-$symmetry
\cite{dorey,most-jordan,GSZ-squire,GS-cjp-2005,MZ-PLB2007}.

Other, non-quantum mechanical examples where EPs play an important
role are the optics of bianisotropic crystals \cite{berry-optics1},
acoustic models \cite{acoust}, many hydrodynamic setups where EPs
have been studied within pseudo-spectral approaches
\cite{pseudo-spec} as well as a large number of mechanical models
\cite{Seyranian-Mailyb-book} where they are connected with regimes
of critical stability \cite{oleg}. Recent results on
magnetohydrodynamic dynamo models indicate on a close connection
between  nonlinear polarity reversal mechanisms of magnetic fields
and EPs \cite{reversal}.

For completeness we note that the perturbation theory for systems in
the vicinity of EPs dates back to 1960 \cite{vishik} (see also
\cite{baumg}) and that it is closely related to singularity theory,
catastrophe theory  and versal deformations of Jordan structures
\cite{arnold-sing1}. Supersymmetric mappings between EP
configurations have been recently considered in
\cite{BFS-JPA-Lett-2005,andrianov-jordan-susy}.

A correct perturbative treatment of models in the vicinity of EPs
has to be built on an expansion in terms of root vectors
(eigenvectors and associated vectors  $\Theta_i(\bX_c)$)  at the
corresponding unperturbed eigenvalue $\lb(\bX_c)$ (see e.g.
\cite{baumg,Seyranian-Mailyb-book}). For $\bX\not\in\cV$ (away from
the EP at $\bX_c$ and from other EPs) the operator $H(\bX)$ has a
diagonal spectral decomposition with corresponding eigenvectors
$\Phi_i(\bX)$. Choosing the normalization of these eigenvectors away
from the EP and without regard to the expansion in terms of root
vectors leads to a divergence of the normalization constants in the
EP-limit $\bX\to\bX_c$. The diagonalization break-down at $\bX_c$,
the occurrence of the Jordan block structure and the singular
behavior of the eigenvector decomposition are generic and were many
times described in various contexts (see e.g.
\cite{moiseyev,andrianov-jordan-jpa}). The natural question
connected with the fitting of the root-vector based normalization
and the diagonalizable-configuration normalization (and related
controversial discussions on their physical interpretation
\cite{moiseyev,andrianov-jordan-jpa}) is whether and how the
singular behavior affects the projective Hilbert space structure of
quantum systems.

In the present paper we answer this question by resolving the
singularity with the help of embedding the original Hilbert space
$\cH=\CC^n$ into its projective extension $\CC\PP^n$ instead of
projecting it down to $\CC\PP^{n-1}$ as in standard Hermitian
Quantum Mechanics. Diagonal spectral decompositions and
decompositions with nontrivial root spaces live then simply in
different (and complementary) affine charts of $\CC\PP^n$ similar
like monopole configurations of Hermitian systems have to be covered
with two charts (North-pole chart and South-pole chart) of the unit
sphere $S^2$ \cite{nakahara}.

The basic construction is demonstrated on a simple complex symmetric
(non-Hermitian) $2\times 2-$matrix toy model. The consideration of
complex symmetric matrices sets no restriction because by a
similarity transformation any complex matrix can  be brought to a
complex symmetric form (see, e.g. \cite{gantmacher,braendas}). The
Hilbert space notations for the $2\times 2-$matrix model are fixed
in section \ref{setup}. In section \ref{jordan}, following
\cite{MKS-1,Seyranian-Mailyb-book} we derive the leading-order
perturbative expansion in the vicinity of an EP at $\bX_c$ in terms
of root vectors and fit it then explicitly with expressions of the
diagonal spectral decomposition at $\bX\neq \bX_c$. Combining
geometric phase techniques for non-Hermitian systems \cite{garrison}
with projective Hilbert space techniques from \cite{page-1} we
generalize the projective geometric phase techniques of Hermitian
systems to paths around EPs (section \ref{geom}). The corresponding
monodromy group is identified as parabolic Abelian subgroup of the
special linear group $SL(2,\CC)$ and evidence is given that vector norm
scalings are only due to complex dynamical phases whereas
geometrical phases are purely real-valued and norm preserving. In
section \ref{instant} we consider an instantaneous (stationary type)
picture of the system. Within such a picture, we resolve the
singular normalization behavior by projectively embedding the
Hilbert space $\cH=\CC^2\hookrightarrow\CC\PP^2$. We discuss the
necessity for an affine multi-chart covering of $\CC\PP^2$ in order
to accommodate diagonal-decomposition vectors and root vectors at
EPs simultaneously. The usefulness of the phase rigidity as distance
measure to EPs is discussed in section \ref{rigid}. In section
\ref{PTSQM} some EP-related aspects of $\cP\cT-$symmetric quantum
models are discussed and a conjecture concerning the quantum
brachistochrone problem \cite{CMB-brachist,fring-brach} is
formulated. The Conclusions \ref{conclu} are followed by
\ref{jordan-derive} where auxiliary results on Jordan structures of
complex symmetric matrices are listed.

\section{Setup\label{setup}}
Subject of our consideration is the behavior of a quantum system
near a level crossing point of two resonance states  --- supposing
that for an $N-$level system the influence of the other $N-2$
levels is sufficiently weak. Under this assumption the setup can
be modeled by an effective complex symmetric (non-Hermitian)
$2\times 2$ matrix Hamiltonian
\be{a1}
H=\left(
\begin{array}{cc}
\epsilon_1 &  \omega \\ \omega & \epsilon_2
\end{array}
\right), \qquad H=H^T .
\ee
The effective energies $\epsilon_{1,2}\in \CC$ and the effective
channel coupling $\omega\in\CC$ will in general depend on
underlying parameters $\bX=(X_1,\ldots,X_k)\in\cM$ from a space
$\cM$.

For nonvanishing coupling $\omega\neq 0$ the Hamiltonian can be
rewritten as
\be{a1a}
H=E_0\otimes I_2 + \omega \left(
\begin{array}{cc}
Z &  1 \\ 1 & -Z
\end{array}
\right)
\ee
with $I_2$ denoting the $2\times 2$ unit matrix and
\be{a4}
E_0:= \frac 12 (\epsilon_1+\epsilon_2),\qquad
Z:=\frac{\epsilon_1-\epsilon_2}{2\omega}\,.
\ee
In this representation the eigenvalues $E_\pm$ and eigenvectors
$\Phi_\pm$ of $H$ take the very simple form
\be{a3}
E_\pm=E_0\pm  \omega \sqrt{Z^2+1}
\ee
and
\be{a8}
\Phi_\pm=\left(\begin{array}{c}1\\
-Z\pm \sqrt{Z^2+1}\end{array}\right)c_\pm\,, \quad c_\pm \in
\CC^*:=\CC-\{0\}
\ee
which makes the branching behavior most transparent\footnote{The
fact that $\Phi_\pm$ depends only on the single parameter $Z$
reflects the property that after rescaling the energy by
$1/\omega$ and shifting it by $-E_0/\omega$ (these transformations
do not affect the eigenvectors) the Hamiltonian \rf{a1a} depends
only on $Z$.}. From the overlap
\ba{a3a}
&& \langle\Phi_+|\Phi_-\rangle\equiv \Phi^{*T}_+ \Phi_-\\
&=&c_+^*c_-\left[1+|Z|^2-|Z^2+1|+2 \Im
\left(Z^*\sqrt{Z^2+1}\right)\right]\nn
\ea
one reads off that $\langle\Phi_+|\Phi_-\rangle= 0$ holds only for
$\Im Z= 0$ and that for general $Z\in \CC$ the two states $\Phi_+$
and $\Phi_-$ are not orthogonal $\langle\Phi_+|\Phi_-\rangle\neq 0$.
Following standard techniques \cite{Seyranian-Mailyb-book} for
non-Hermitian operators, we consider a dual (left) basis $\Xi_\pm$
bi-orthogonal to $\Phi_\pm$
\be{a3b}
(H^+-E_\pm^*)\Xi_\pm =0, \qquad \langle\Xi_k|\Phi_l\rangle\propto
\delta_{kl},\quad k,l=\pm\,.
\ee
For complex symmetric $H$ it holds $\Xi_\pm\propto \Phi_\pm^*$ so
that the most general ansatz for the right and left basis vectors
$\Phi_\pm$ and $\Xi_\pm$ can be chosen as
\ba{a3c} &&\Phi_\pm=c_\pm \chi_\pm,\quad
\Xi_\pm=d_\pm^*\chi_\pm^*,\quad c_\pm,d_\pm\in\CC^*\label{a3c-1}\\
&&\chi_\pm:=\left(\begin{array}{c}1\\
-Z\pm \sqrt{Z^2+1}\end{array}\right)\label{a3c-2}\,.
\ea
The bi-orthogonality
\be{i1}
\langle\Xi_\pm|\Phi_\mp\rangle=d_\pm c_\mp \chi_\pm^T \chi_\mp=0
\ee
is ensured by the structure of $\chi_\pm$ and holds for any value
of the parameter $Z\in \CC$. A normalization
$\langle\Xi_\pm|\Phi_\pm\rangle=1$ would set two constraints on
the four free scaling parameters $c_\pm,d_\pm\in\CC^*$
\be{i1a}
\langle\Xi_\pm|\Phi_\pm\rangle=d_\pm c_\pm \chi_\pm^T\chi_\pm=1\,,
\ee
so that the system would still have two free parameters which
should be fixed by additional assumptions. Subsequently, we will
mainly work with the bi-orthogonality properties of the vectors
$\Phi_\pm$, $\Xi_\pm$ and fix their normalization only when
explicitly required.

Due to the arbitrary scaling parameters $c_\pm,d_\pm \in \CC^*$ of
the right and left eigenvectors $\Phi_\pm,\Xi_\pm \in \cH=\CC^2$
\rf{a3c-1} it is natural to consider equivalence classes of such
vectors defined by corresponding lines $\pi(\Phi_\pm )$,
$\pi(\Xi_\pm ) $. These lines form the projective Hilbert space
$\PP(\cH)=\cH^*/\CC^*=\CC\PP^1\ni \pi(\Phi_\pm),\pi(\Xi_\pm )$
\cite{kodaira,mumford,naber}, where $\cH^*:=\cH-\{0\}$ denotes the
original Hilbert space with the point at origin $\{0\}=(0,0)$
deleted to allow for a consistent definition of $\PP(\cH)$. The
space $\PP(\cH)$ is covered by a single chart of homogeneous
coordinates $(z_0,z_1)^T\in \cH$ and two complementary charts of
affine coordinates $U_0\ni (1,z_1/z_0),\ z_0\neq 0$ and $U_1\ni
(z_0/z_1,1),\ z_1\neq 0$. Comparison with the structure of the
auxiliary vectors $\chi_\pm$ \rf{a3c-2}  shows that the $\chi_\pm$
can be straight forwardly re-interpreted as  points of the
projective space $\CC\PP^1$ described by the affine coordinate over
$U_0$: \ \ $\chi_\pm\approx \pi(\Phi_\pm)$.  The vectors
$\{\Phi_\pm,\Xi_\pm\}$ themselves can be understood as sections of
the natural line bundle $L=\{(p,v)\in\PP(\cH)\times \cH|\, v=cp,\,
c\in \CC^*\}$ \cite{nakahara}, i.e. as
$\Phi_\pm=\pi(\Phi_\pm)\otimes c_\pm$, $\Xi_\pm=\pi(\Xi_\pm)\otimes
d^*_\pm$, where $\pi$ denotes the projection $\pi: \ \cH^* \to
\PP(\cH)$. The bundle structure is locally trivial
$\pi^{-1}(U_0)\approx U_0\times \CC^* \ni \Phi_\pm$
\cite{kobayashi}\footnote{For completeness we note that the (right)
eigenvectors $\Phi_\pm$ and the dual (left) ones $\Xi_\pm$ could be
understood as elements of a vector bundle $\PP(\cH)\times F$ and its
dual $\PP(\cH)\times F^*$ with pairing in the fibres
$\langle.|.\rangle:\ F_p^*\times F_p\longrightarrow \CC$ (see, e.g.
\cite{kodaira}). The details of this construction will be presented
elsewhere.}.

\section{Jordan structure \label{jordan}}

At an EP, the two eigenvalues $E_\pm$ coalesce $E_+=E_-=E_0$.
According to (\ref{a3}), this happens for $Z^2=-1$ and $Z=Z_c:=\pm
i$ and via \rf{a3c-1} it is connected with a coalescence of  the
corresponding lines $\pi(\Phi_+)=\pi(\Phi_-)=:\pi(\Phi_0)$ encoded
in
\be{i2}
\chi_+=\chi_-=\chi_0:=\left(\begin{array}{c}1\\
-Z_c\end{array}\right)=\left(\begin{array}{c}1\\
\mp i\end{array}\right).
\ee
This means that the eigenvalue $E_0$ has algebraic multiplicity
$n_a(E_0)=2$ and geometric multiplicity $n_g(E_0)=1$ and by
definition the level crossing point is an EP of the spectrum. The
bi-orthogonality \rf{i1} of $\Phi_\pm$ and $\Xi_\mp$ is compatible
with the coalescence of the lines due to the vanishing bi-norm
$\chi_0^T\chi_0=0$, i.e. the isotropy\footnote{The vector $\chi_0$
behaves similar like a vector on the light cone in Minkowski
space.} of $\chi_0$,
--- a generic fact holding for the (geometric) eigenvector at any
EP \cite{baumg,Seyranian-Mailyb-book}. We note that the coalescence
$\pi(\Phi_+)=\pi(\Phi_-)=\pi(\Phi_0)$ still leaves the freedom for
the vectors $\Phi_+=c_+\chi_0$ and $\Phi_-=c_-\chi_0$ of being two
different sections $\Phi_+\neq \Phi_-$ of the same fiber
$\pi(\Phi_0)\times\CC^*$ over $\pi(\Phi_0)$.

The right and left eigenvectors $\Phi_0$, $\Xi_0$ at the EP are
supplemented by corresponding associated vectors (algebraic
eigenvectors) $\Phi_1$ and $\Xi_1$ defined by the Jordan chain
relations \cite{Seyranian-Mailyb-book}
\ba{i3}
[H(Z_c)-E_0I_2]\Phi_0=0,&  &[H(Z_c)-E_0I_2]\Phi_1=\Phi_0\\
\left[H(Z_c)-E_0I_2 \right]^+ \Xi_0=0,& &\left[H(Z_c)-E_0I_2
\right]^+ \Xi_1=\Xi_0\,.\nn
\ea
From the inhomogeneity of these Jordan chain relations it follows
immediately that the root vectors $\Phi_0$ and $\Phi_1$ as well as
$\Xi_0$ and $\Xi_1$ scale simultaneously and in a linked way with
the same {\em single} scale factor $c_0$ and $d_0^*$, respectively.
This is also visible from their explicit representation  \rf{qa8}
derived in  \ref{jordan-derive}
\ba{i4}
\Phi_0=\sigma qc_0\left(%
\begin{array}{c}
  1 \\
  -Z_c \\
\end{array}%
\right), && \Phi_1=\sigma q^{-1}c_0\left(%
\begin{array}{c}
    -Z_c \\
    1 \\
\end{array}%
\right)\nn\\
\Xi_0=\sigma q^*d_0^*\left(%
\begin{array}{c}
    -Z_c \\
    1 \\
\end{array}%
\right), && \Xi_1=\sigma q^{*-1}d_0^*\left(%
\begin{array}{c}
        1 \\
        -Z_c \\
\end{array}%
\right)\\
\sigma:=\frac{e^{i\mu\frac \pi 4}}{\sqrt 2},\ q:=\sqrt{2\omega}, &&
Z_c=\pm i=:\mu i, \ \  c_0,d_0\in \CC^*\,.
\ea
The simultaneous scaling means that the lines $\pi(\Phi_0)$,
$\pi(\Xi_0)$ at the EP should be interpreted as the one-dimensional
components (projections) of two-dimensional planes which span the
root space\footnote{In the present simplest model $\fS(E_0)$ fills
the whole Hilbert space $\cH=\CC^2$.} $\fS(E_0)$ \cite{baumg} and
which scale as a whole with a single scale factor. Such a
higher-dimensional (complex) plane-structure goes clearly beyond the
one-dimensional line structure of the projective space $\PP(\cH)$
(mathematically one should extend the natural line bundle of the
original projective space to a more general projective flag bundle
\cite{encyc-jap1,parabolic-flag-book})\footnote{Indications that all
the root vectors of a Jordan chain should scale simultaneously with
a single scale factor were given, e.g., in \cite{bohm-jordan} for
Gamow vector setups with higher S-matrix poles.} and underlines the
fact that the state at an EP itself is not an element of the
projective Hilbert space $\PP(\cH)$ in its usual understanding.

The basis sets $\{\Phi_0,\Phi_1\}$ and $\{\Xi_0,\Xi_1\}$ satisfy
the well known bi-orthogonality conditions
\cite{Seyranian-Mailyb-book}
\ba{i5}
&&\langle\Xi_0|\Phi_0\rangle=\langle\Xi_1|\Phi_1\rangle=0\nn\\
&&\langle\Xi_0|\Phi_1\rangle=\langle\Xi_1|\Phi_0\rangle=d_0c_0\neq
0\,.
\ea
Again, a normalization
$\langle\Xi_0|\Phi_1\rangle=\langle\Xi_1|\Phi_0\rangle=1$ would
only lead to a constraint $d_0c_0=1$ on the scale factors, but
would not fix them completely. Due to this scaling freedom the
single line $\pi(\Phi_0)$ of a given Jordan structure, in general,
still allows for different sections $\Phi_{0,a}\neq\Phi_{0,b}$ of
the corresponding fiber $\pi(\Phi_0)\times \CC^*\ni
\Phi_{0,a},\Phi_{0,b}$,
$\pi(\Phi_{0,a})=\pi(\Phi_{0,b})=\pi(\Phi_0)$.

Let us now consider in detail what happens when the system
approaches one of the critical values $Z_c=\pm i$. For this purpose
we use the well-defined (but completely general and arbitrary)
ansatz
\be{a14}
Z=Z_c+\varepsilon,\qquad |\varepsilon|\ll 1, \qquad \varepsilon\in
\CC
\ee
and expand the eigenvalues \rf{a3} and the line defining vectors
$\chi_\pm$ \rf{a3c-2} in terms of $\varepsilon$. This gives the
leading contributions to their Puiseux series representation
\cite{baumg,Seyranian-Mailyb-book} in $\varepsilon^{1/2}$ as
\ba{i6} E_\pm &=&E_0\pm \varepsilon^{1/2}\Delta
E+o(\varepsilon^{1/2}), \label{i6a}\\
\Delta E&:=&\omega \sqrt{2Z_c}\,,\nn\\
 \chi_\pm &=&\left(
           \begin{array}{c}
             1 \\
             -Z_c \\
           \end{array}
         \right)\pm \varepsilon^{1/2}\left(
           \begin{array}{c}
             0 \\
             \sqrt{2Z_c} \\
           \end{array}
         \right)+o(\varepsilon^{1/2})\label{i6b}\,.
\ea
Following \cite{MKS-1,Seyranian-Mailyb-book}, we expand the
eigenvectors $\Phi_\pm (Z)$ of the diagonal spectral decomposition
in the same local $\varepsilon^{1/2}$ approximation in terms of
the Jordan chain (root) vectors $\Phi_{0,1}$
\ba{i7}
\Phi_\pm
&=&\Phi_0+\varepsilon^{1/2}(b_0\Phi_0+b_1\Phi_1)+o(\varepsilon^{1/2}),\label{i7a}\\
b_0&=&\pm\frac{Z_c}{2\omega}\Delta E,\qquad b_1=\pm \Delta E\,.
\label{i7b}
\ea
The coefficients $b_{0,1}$ are obtained by a two-step procedure.
Substituting \rf{a14}, \rf{i6a}, \rf{i7a} into the eigenvalue
equation and making explicit use of the chain relations \rf{i3}
yields $b_1$ and leaves $b_0$ still undefined. The coefficient $b_0$
is found by comparing the line structures\footnote{The term
$\varepsilon^{1/2}b_0\Phi_0$ additionally present in \rf{i7a} in
comparison with the corresponding result in \cite{MKS-1} is due to
the different choice of the root (Jordan chain) vectors $\Phi_0$,
$\Phi_1$. The chain relation \rf{i3} shows that the associated
vector $\Phi_1$ is defined up to additional $\Phi_0$ contributions
and can be replaced by any linear combination $\Phi_1+a\Phi_0, \ \
a\in \CC$.} of $\Phi_\pm$ with $\chi_\pm$ in \rf{i6b}.

It remains to match the fiber sections --- what can be done in two
ways. One may assume a single scaling coefficient $c_0$ of the
root space given and consider the coefficients $c_\pm$ of the
sections $\Phi_\pm$ as derived objects. This leads to the
identification $c_+=c_-=\sigma q c_{0}.$ Apart from this option,
one may assume the scaling coefficients $c_\pm$ as primary objects
and given so that they may take different values $c_+\neq c_-$.
Correspondingly the scaling factor $c_0$ of the root space will
then be fitted to $c_\pm$ so that it will take two different
values
\be{i8}
c_{0,\pm}=c_\pm/(\sigma q)\,.
\ee
Both constructions are possible and compatible with the smooth
fitting of the line structure encoded in the EP-limiting behavior
$\pi(\Phi_\pm)\to \pi(\Phi_0)$.

In a way similar to the above two-step procedure with subsequent
fiber fitting the left eigenvectors can be obtained as
\ba{i9}
\Xi_\pm^*&=& \Xi_0^*+\varepsilon^{1/2}(b_0\Xi_0^*+b_1\Xi_1^*)+o(\varepsilon^{1/2}),\label{i8a}\\
d_\pm&=&\sigma^*qZ_c d_{0,\pm}\,.\label{i8b}
\ea
Here, $b_0$ and $b_1$ are the same as in \rf{i7b} and full
compatibility with the bi-orthogonality conditions \rf{a3b} as well
as with \rf{a3c-1} is easily verified by direct calculation. In case
of a single scaling factor $d_0$ of the dual root space the
coefficients $d_\pm$ will coincide $d_+=d_-=\sigma^*qZ_c d_0$.

Combining \rf{i7} and \rf{i9} one finds the limiting behavior of the
inner products as
\ba{i10}
\langle\Xi_\pm|\Phi_\pm\rangle&=&2b_1d_{0,\pm}c_{0,\pm}\varepsilon^{1/2}+
o(\varepsilon^{1/2})\nn\\
&=&\frac{2b_1}{\omega Z_c}d_\pm c_\pm \varepsilon^{1/2}+
o(\varepsilon^{1/2})\,.
\ea
Here, one has to distinguish two normalization schemes. If one
assumes the root vector sets $\{\Phi_0,\Phi_1\}$, $\{\Xi_0,\Xi_1\}$
normalized, e.g., with $d_{0,\pm}c_{0,\pm}=1$ or $d_0c_0=1$ in
\rf{i5} then the scalar product $\langle\Xi_\pm|\Phi_\pm\rangle$ of
the eigenvectors in the diagonal spectral decomposition (see
\rf{i10}) vanishes in the EP-limit. Starting,  in contrast, from
normalized eigenvector pairs $\{\Phi_\pm,\Xi_\pm\}$ of
diagonalizable Hamiltonians as in \rf{i1a}, i.e. with
$\langle\Xi_\pm|\Phi_\pm\rangle=1$, then the scale factor products
$d_\pm c_\pm$ diverge as $d_\pm c_\pm\propto \varepsilon^{-1/2}$ for
$\varepsilon\to 0$. Both normalization schemes are possible and
compatible with the smooth limiting behavior $\pi(\Phi_\pm)\to \pi
(\Phi_0)$ of the lines encoded in $\chi_\pm(\varepsilon\to 0)\to
\chi_0$ [cf. \rf{i2}]. We see that this special behavior is only
related to the fiber sections and not to the fibers (lines)
themselves. The two incompatible normalization schemes simply
indicate on the need for two complementary charts to cover the whole
physical picture in the vicinity of a $2\times 2$ Jordan block
$J_2$.  One of these charts (we will call it the root vector chart)
remains regular in the EP-limit, whereas the other (diagonal
representation) chart becomes singular.

The situation is similar to the two affine charts required to
cover the Riemann sphere $\CC\PP^1$. Starting from homogeneous
coordinates $(x_0,x_1)\in \CC\PP^1$ one has the two affine charts
$U_0\ni(1,x_1/x_0)$, $x_0\neq 0$ and $U_1\ni(x_0/x_1,1)$, $x_1\neq
0$. The mutually complementary affine coordinates $z:=x_1/x_0\in
\CC^1$ and $w:=x_0/x_1\in \CC^1$ are then related by the well
known fractional transformation $w=1/z$ so that the singular
$|z|\to \infty$ limit in the $z-$chart corresponds simply to the
regular $w\to 0$ limit in the $w-$chart. In other words, the two
charts cover the North-pole region and the South-pole region of
the Riemann sphere --- a construction well known, e.g., from
complex analysis and the description of magnetic monopoles
\cite{nakahara}.

Returning to the two-chart picture of the normalization we see that
the original Hilbert space $\cH=\CC^2$ should be extended by the set
of infinite vectors $\Phi_\pm$ what can be naturally accomplished by
embedding it into a larger projective space
$\cH\hookrightarrow\CC\PP^2$. Correspondingly the fibers
$\pi(\Phi_\pm)\times \CC^*$ should be extended as
$\pi(\Phi_\pm)\times \CC^*\hookrightarrow
\pi(\Phi_\pm)\times\CC\PP^1$. A detailed discussion of these
structures will be presented in \cite{G-new1}. An explicit embedding
construction for simplified setups with coinciding scale factors
$d_\pm=c_\pm$ is given in section \ref{instant} below.

\section{Geometric phase\label{geom}}
Following earlier studies
\cite{heiss-harney-chirality,garrison,berry-naples,berry-non-herm1,mondrag-geom},
geometric phases \cite{berry-phase0} of eigenvectors of
non-Hermitian complex symmetric operators  have been recently
considered in \cite{MKS-1} for paths in parameter space encircling
an EP. The results showed full agreement with the phase
considerations of \cite{heiss-harney-chirality}. In this section,
we combine techniques for non-Hermitian systems
\cite{garrison,MKS-1} with explicit projective space
parameterizations for Hermitian systems \cite{page-1} to provide
an explicit projective-space based derivation of the phase
representation for non-Hermitian systems. Such an explicit
reshaping of the results of \cite{page-1} to non-Hermitian setups
seems missing up to now.

Following \cite{garrison,berry-naples,berry-non-herm1,MKS-1} we
consider an auxiliary system with a general non-Hermitian
Hamiltonian $H(t)$ depending on a set of non-stationary parameters
$\bX(t)=[X_1(t),\ldots, X_m(t)]\in \cM$, $H(t)=H[\bX(t)]$ and an
EP hyper-surface $\cV\subset\cM$ which is encircled by an
appropriate loop $\Gamma$ in parameter space
$\cM\ni\Gamma=\{\bX(t),\ t\in[0,T]:\ \bX(0)=\bX(T)\}$. The
evolution of the quantum system is governed by a usual
Schr\"odinger equation for the right eigenvectors
\be{g1}
i\p_t\Phi_n(t)=H(t)\Phi_n(t),
\ee
and, due to the time invariance of the bi-orthogonal product
\be{g2}
\langle\Xi_m(t)|\Phi_n(t)\rangle=\d_{mn}\,,
\ee
by a complementary evolution law for the left eigenvectors
\cite{garrison}
\be{g3}
i\p_t\Xi_m(t)=H^+(t)\Xi_m(t)\,.
\ee
For an adiabatic motion cycle $\Gamma\subset\cM$ with Hamiltonian
$H[\bX(T)]=H[\bX(0)]$ the resulting eigenvector $\Phi_n(t=T)$ of
$H[\bX(T)]$ must lay on the same line as the initial $\Phi_n(t=0)$,
i.e. it can only obtain an additional scaling factor which we
parameterize as complex-valued phase
\be{g4}\Phi_n(T)=e^{i\phi_n(T)}\Phi_n(0),\quad \phi_n(T)\in\CC.
\ee
Due to the preserved orthonormality \rf{g2} the corresponding left
eigenvectors evolve as
\be{g5}
\Xi_m(T)=e^{i\phi_m^*(T)}\Xi_m(0)\,.
\ee
The complex phase $\phi_n(T)$ can be split into a dynamical
component \cite{garrison,page-1}
\be{g6}
\epsilon_n(T)=-\int_0^T\frac{\langle\Xi_n(t)|H(t)|\Phi_n(t)\rangle}{\langle\Xi_n(t)|\Phi_n(t)\rangle}dt
\ee
and the geometric phase
\be{g7}
\gamma_n(T)=\phi_n(T)-\epsilon_n(T)\,.
\ee
Adapting the techniques of \cite{page-1} we calculate $\gamma_n(T)$
in terms of explicit projective space coordinates. Setting
\ba{g8}
\Phi_n(t)&=&c_n(t)\chi_n(t)
=:[z_0(t),z_1(t)]^T=z_0(t)[1,w(t)]^T\nn\\
\Xi_m(t)&=&d_m^*(t)\chi_m^*(t) =:[y_0(t),y_1(t)]^T=y_0(t)[1,v(t)]^T
\ea
(omitting in the projective space coordinates the mode indices
$m$, $n$) one identifies
\be{g9}
\phi_n(T)=-i\ln \left[z_0(T)/z_0(0)\right]
\ee
and obtains from \rf{g1}, \rf{g6} and \rf{g7} the differential
1-form of the geometric phase as
\ba{g10}
d\g&=&-i\frac{dz_0}{z_0}+i\frac{y_0^*dz_0+y_1^*dz_1}{y_0^*z_0+y_1^*z_1}
=i\frac{v^*dw}{1+v^*w^*}\,.
\ea
Due to the symmetry \rf{a3c-1} between left and right eigen-lines
this simplifies to
\ba{g11}
d\g=i\frac{\chi^Td\chi}{\chi^T\chi}=\frac i2\, d\ln (1+w^2)\,.
\ea
Similar to results on Hermitian systems \cite{page-1} the
differential 1-form \rf{g10} is independent of the coordinates
$z_0$ and $y_0$ along the fibers and, hence, defines a horizontal
connection over the projective Hilbert space of the system. The
mere difference in the definitions of the projective structures is
in $\CC\PP^1=S^3/S^1$ for Hermitian systems, whereas
$\CC\PP^1=\cH^*/\CC^*$ for non-Hermitian ones \cite{kobayashi}.

Let us now apply the general technique to the concrete $2\times
2-$matrix model \rf{a1}. Parameterizing the cycle around the EP by
\rf{a14} with
\be{g11a}
\varepsilon=r e^{i\a},\qquad \a\in[0,2\pi],\ 0<r\ll 1
\ee
one reproduces the 1-forms of the geometric phases of
\cite{berry-naples}
\be{g12}
d\g_\pm=\frac i4d\ln\varepsilon=-\frac 14 d\a+\frac i4 d\ln r\,.
\ee
In a similar way one obtains the same 1-forms for the corresponding
left eigenvectors $\Xi_\pm$. Upon integration over a full cycle
$\a(T)=\a(0)+2\pi,\ r(T)=r(0)$ one finds
\be{g12a}
\g_\pm(T)-\g_\pm(0)=-\frac\pi 2\,.
\ee

The relation between geometric phases $\g_\pm$ and the cycle phase
$\a$ can be gained also directly from the structure of the sections
$\Phi_\pm$. These sections may be arranged as columns of a
diagonalizable $2\times 2-$matrix
\be{g13}
\bfPhi(\a):=[\Phi_+(\a),\Phi_-(\a)]\,.
\ee
The evolution along a cycle is then encoded in the transformation
matrix $W(\a)=\bfPhi(\a)\left[\bfPhi(0)\right]^{-1}$ which for small
$\varepsilon$ with $0\neq |\varepsilon|\ll 1$ can be calculated from
the representation \rf{i6b} as
\be{g14}
W(\a)=\left[
\begin{array}{cc}
e^{-i\frac \a 4} & 0 \\
2iZ_c \sin\left(\frac \a 4\right) & e^{i\frac \a 4} \\
\end{array}
\right]\,.
\ee
The elements $W(\a)$ form an Abelian parabolic subgroup $P$ of the
special linear group $SL(2,\CC)\supset P$ (see, e.g.,
\cite{encyc-jap1,parabolic-flag-book,parabolic-winternitz})
\be{g15}
W(\a+\b)=W(\a)W(\b)=W(\b)W(\a)
\ee
corresponding to the mapping $e^{i\a}\in S^1\approx U(1)\mapsto
P\subset SL(2,\CC)$. For full cycles $\a=2\pi N,$ $ N\in\ZZ$ they
yield the monodromy transformations \cite{monodromy}
\ba{g16}
&&W_0:=W(0)=I_2,\quad W_1:=W(2\pi)=\left(
                                                   \begin{array}{cc}
                                                     -i & 0 \\
                                                     2iZ_c  & i \\
                                                   \end{array}
                                                 \right),\nn\\
                                                && W_2:=W(4\pi)=W^2(2\pi)=-I_2,\nn\\
                                                 &&
                                                 W_3:=W(6\pi)=-W(2\pi),\quad
                                                 W(8\pi)=I_2=W_0\,.
\ea
The geometric phase \rf{g12} and the monodromy transformations
\rf{g16} show the typical four-fold covering of the mapping
$\a\mapsto \g$ which was earlier described in
\cite{heiss-m-rotter,heiss-01,heiss-harney-chirality,MKS-1} and
experimentally demonstrated in \cite{darmstadt}. A cycle around
the EP in parameter space $\cM$ has to be passed four times in
order to produce one full $2\pi-$cycle in the geometric phase. A
(non-oriented) eigen-line $\pi(\Phi\neq\Phi_0)\in \CC\PP^2$ is
already recovered after two cycles $\pi
(W_2\Phi)=\pi(-\Phi)=\pi(\Phi)$
--- similar to the eigenvalue $E$ which for the $2\times 2-$matrix
lives on a two-sheeted Riemann surface with the same two branch
points  $Z_c=\pm i$ as the line bundle. For the isotropic limiting
vector $\Phi_0$ at the EP it holds (due to \rf{i2})
\be{g16a}
W(\a)\Phi_0=e^{-i\a/4}\Phi_0
\ee
so that the parabolic subgroup $P\ni W(\a)$ can be identified as
invariance group of the projective line at the EP
\be{g16b}
\pi\left(W(\a)\Phi_0\right)=\pi\left(e^{-i\a/4}\Phi_0\right)=\pi(\Phi_0).
\ee

We note that despite the non-Hermiticity of the Hamiltonian $H$ the
geometric phase is purely real
--- as for Hermitian systems. Relations  \rf{g12}, \rf{g12a} show that possible imaginary
phase contributions (which would result in a re-scaling of the
eigenvectors $\Phi_\pm$) are cancelled by the closed-cycle
condition $r(T)=r(0)$. Hence, the non-preservation of the vector
norm in non-Hermitian systems is induced solely by a complex {\em
dynamical} phase $\epsilon$ and  requires the presence of the
bi-orthogonal basis where a decaying behavior of the right
eigenvectors\footnote{For simplicity, we show the relations for
stationary non-Hermitian Hamiltonians $H$ with constant complex
eigenvalues $E_n=\epsilon_n+i\frac{\Gamma_n}2=\const;\ \
\epsilon_n,\Gamma_n\in \RR$.}
\be{g17}
\Phi_n\propto e^{-i\epsilon_n t-\frac{\Gamma_n}2 t},\quad
\langle\Phi_n|\Phi_n\rangle=||\Phi_n||^2\propto e^{-\Gamma_n t}
\ee
is necessarily connected with increasing vector norms of the dual
left eigenvectors
\be{g18}
\Xi_m\propto e^{-i\epsilon_m t+\frac{\Gamma_m}2 t},\quad
||\Xi_m||^2\propto e^{\Gamma_m t}
\ee
so that indeed $\langle\Xi_m|\Phi_n\rangle=\delta_{mn}$. This
behavior is well known from resonances and Gamow vector theory (see,
e.g. \cite{mondragon-1}).

Comparison of \rf{a3c-1} and \rf{g17}, \rf{g18} shows that the
formal ansatz $\Xi_m=\Phi_m^*$ for the eigenvectors of the complex
symmetric Hamiltonian (cf. \cite{marost23,rs1,rs2}) can be used only
for instantaneous eigenvectors at a single fixed $t=t_0$ (which
formally can be set to $t_0=0$) as well as for the subclass of real
symmetric matrices (when the system becomes Hermitian and
norm-preservation of the eigenvectors holds). In contrast, for
explicitly time dependent non-Hermitian setups it only holds
$\Xi_m(t)\propto\Phi_m^*(t)$, i.e. the dual basis vectors
necessarily live on complex conjugate lines (fibers) $\pi
[\Xi_m(t)]=\left(\pi [\Phi_m(t)]\right)^*$ but with
$\Xi_m(t)\neq\Phi_m^*(t)$ for $t\neq t_0$.

Aspects of the parameter dependence  of the phases and scalings in
an instantaneous picture with $\Xi_m=\Phi_m^*$  together with the
explicit EP-limit $\varepsilon\to 0$ are subject of the next
section.

\section{Instantaneous picture\label{instant}}

In modern quantum physics not only the properties of natural
systems such as nuclei or atoms are of interest, but rather the
design and functionality of artificial quantum-system-based
devices plays an essential role. In many cases, for the
understanding of the dynamical features of these man-made quantum
systems the time dependence is of minor interest. The properties
of these systems are mainly governed by the position and number of
EPs, i.e. the level crossing points in the complex plane, and
their dependence on external control parameters. In this context
it appears natural to study the parameter dependence of level
energies and widths as well as the corresponding eigenvectors in
terms of the instantaneous picture with $\Xi_m=\Phi_m^*$ and
$c_\pm=d_\pm$. This picture is compatible with the Hermitian limit
when $\Im \epsilon_{1,2}=\Im \omega=0$ in \rf{a1} and the
condition $\Xi_m=\Phi_m^*$ is fulfilled by
definition\footnote{When compatibility with the Hermitian limit is
not required, then the bi-orthonormalization constraints
$d_{0,\pm}c_{0,\pm}=1$ or $d_\pm c_\pm\chi_\pm^T\chi_\pm=1$ fix
only two of the four constants $c_{0,\pm},d_{0,\pm}$ or $c_\pm,
d_\pm$ and the remaining two can be chosen arbitrarily. For
instance, one may set the eigenvector scaling factors as
$c_{0,\pm}=C\neq 1$ or $c_\pm=1$ so that $d_{0,\pm}=C^{-1}$ or
$d_\pm=\left(\chi_\pm^T\chi_\pm\right)^{-1}$ what would define
instantaneous pictures not compatible with the Hermitian limit.}.

We have to distinguish the two possible normalization schemes ---
the root-vector based normalization \rf{i5} with $d_0c_0=1$ or
$d_{0,\pm}c_{0,\pm}=1$ and the diagonal-representation based
normalization \rf{i1a} with $d_\pm c_\pm \chi^T\chi=1$.

In the root-vector based normalization scheme the conditions
$c_\pm=d_\pm$ and $d_{0,\pm}c_{0,\pm}=1$ together with the two
relations \rf{i8} and \rf{i8b} imply (in leading-order approximation
in $\varepsilon$) $c_{0,\pm}=d_{0,\pm}$ and, hence,
$c_{0,\pm}=\kappa$ with $\kappa =\pm 1$ (independently of the signs
$\pm $ in the index of $c_{0,\pm}$) as well as $c_\pm=d_\pm =\kappa
\sigma q$. We see that in leading-order approximation in
$\varepsilon$ the scaling factors $c_\pm=d_\pm$ are rigidly fixed
and independent of $\varepsilon$. A geometric phase (necessarily
induced via an $\varepsilon-$dependence) appears incompatible with
this normalization.

Let us now turn to the diagonal-representation based normalization
\rf{i1a}. In the EP-limit $\e\to 0$ the normalization condition
\rf{i1a} for the eigenvectors \rf{a8} yields
\ba{r2}
1=\langle\Xi_\pm|\Phi_\pm\rangle=\Phi^T_\pm \Phi_\pm
&=&\left[1+\left(Z\mp
\sqrt{Z^2+1}\right)^2\right]c^2_\pm\nonumber\\
&\approx&\mp 2Z_c\sqrt{2Z_c\varepsilon}\, c^2_\pm
\ea
and we find the expected divergent scaling factors as
\ba{r3}
c^2_\pm\approx \mp 2^{-3/2}Z_c^{-3/2}\varepsilon^{-1/2}\qquad
\Longrightarrow \qquad  c_\pm \sim \varepsilon^{-1/4}\,.
\ea

On the one hand, \rf{r3} reproduces the local four-sheeted Riemann
surface structure connected with the geometric phase \rf{g12},
\rf{g14}, i.e. a fourfold winding around the EP is needed to
return to an eigenvector pointing into the same complex direction
as a starting vector. (In contrast to the root-vector
normalization scheme full compatibility with the geometric phase
setup holds.)

On the other hand, it leads to divergent vector norms
\be{r6}
||\Phi_\pm||^2=\langle\Phi_\pm|\Phi_\pm\rangle\approx
2|c_\pm|^2\approx |2\varepsilon|^{-1/2}
\ee
for $\varepsilon\to 0$. As it was indicated in section \ref{jordan},
the corresponding singularity can be naturally resolved by embedding
the original Hilbert space $\cH\approx \CC^2$ into its projective
extension $\cH\hookrightarrow \CC\PP^2\ni \phi=(u_0,u_1,u_2)$ so
that the set of infinite vectors becomes well defined. Interpreting
the two components $z_0$ and $z_1$ of the vector (fiber section)
\be{r7}
\Phi=c(1,w)=(z_0,z_1)\in \CC^2
\ee
as affine coordinates on the chart
$U_2\ni(\frac{u_0}{u_2},\frac{u_1}{u_2},1)$, $u_2\neq 0$,
$U_2\subset \CC\PP^2$
\be{r8}
\Phi=(c,cw)\hookrightarrow (c,cw,1)
\ee
we can identify $\Phi$ with the point $\phi\in\CC\PP^2$ with
homogeneous coordinates
\be{r9}
\phi=(u_0,u_1,u_2)=(1,w,c^{-1}).
\ee
The singularity $|c|\to\infty$ at the EP corresponds then simply
to the point $\phi_0=(1,w,0)\in \CC\PP^2$ with $u_2=0$ and we see
that the affine chart $U_2\in\CC\PP^2$ is no longer appropriate
for covering  $\phi_0$. This is in contrast to the root-vector
normalization scheme where $c$ is fixed and the chart $U_2$
remains suitable for the covering. Within the present
diagonal-representation normalization,  instead, $\phi_0$ should
be parameterized in terms of affine coordinates corresponding  to
one of the charts\footnote{A projective space $\CC\PP^n\ni
(z_0,z_1,\ldots,z_n )$ is covered by $n+1$ affine charts $U_k\ni
(\frac{z_0}{z_k},\ldots,
\frac{z_{k-1}}{z_k},1,\frac{z_{k+1}}{z_k},\ldots,\frac{z_n}{z_k})$
with $z_k\neq 0$ (see, e.g., \cite{mumford,kobayashi}) in straight
forward dimensional extension of the two-chart covering of the
Riemann sphere $\CC\PP^1$ mentioned in section \ref{jordan}.}
$U_0$ or $U_1$ with $u_0\neq 0$ or $u_1\neq 0$. Most natural for
our representation \rf{r7}, \rf{r9} is the affine chart $U_0\ni
(1,\frac{u_1}{u_0},\frac{u_2}{u_0})$ which can be used for a
suitable projective representation of the fibre sections $\Phi$
\be{r9a}
\Phi\approx (1,w,c^{-1})=(\chi^T,c^{-1})\approx
(\pi(\Phi),c^{-1})\,.
\ee

Interpreting the normalization condition \rf{r2} as constraint on
the affine coordinates of $\Phi$ in the chart $U_2$
\be{r10}
0=\Phi^T\Phi-1=\frac{u_0^2}{u_2^2}+\frac{u_1^2}{u_2^2}-1
\ee
one immediately sees that it is equivalent to the conic (singular
quadric)\footnote{For conics and quadrics in projective spaces
see, e.g. \cite{mumford,encyc-jap1,candelas-ossa}.}
\be{r11}
u_0^2+u_1^2-u_2^2=0
\ee
in homogeneous coordinates which cover the whole $\CC\PP^2$. This
conic remains regular at EPs which merely correspond to
configurations with $u_2=0$.  In terms of
$(\chi^T,c^{-1})-$notations it reads
\be{r12}
\chi^T\chi-c^{-2}=0\,.
\ee
It is clear that the conic construction is straight forwardly
extendable to Hilbert space embeddings $\cH=\CC^n\hookrightarrow
\CC\PP^n$ of any dimension $n$. We arrive at the conclusion that the
appropriate state space for open quantum systems in an instantaneous
setting will be related to the projective extension $\CC\PP^n$ of
the original Hilbert space $\cH=\CC^n$ with states identified with
conics $\sum_{k=0}^{n-1}u_k^2-u_n^2=0$. This is in contrast to
Hermitian systems where it is sufficient to project the Hilbert
space $\cH^*=\CC^n-\{0\}$ down to the base space $\CC\PP^{n-1}$,
i.e. $\pi:\ \cH^*\to \PP(\cH^*)\approx \CC\PP^{n-1}$. In
non-Hermitian setups each fiber $\pi(\Phi)\times \CC^*$ should be
supplemented by $\infty$. This suggests to extend them to
$\pi(\Phi)\times\CC\PP^1$. From the above construction we see that
the singular behavior with regard to the two affine charts is only
related to the scale factors $c\in\CC\PP^1$, whereas $\pi(\Phi)$
behaves smoothly and regular. On its turn, this suggests to
reconsider the model dependent physical interpretation of the
eigenvector self-orthogonality (isotropy) and the corresponding
diverging or non-diverging sensitivity in perturbation expansions
like in \cite{moiseyev,andrianov-jordan-jpa} as result of divergent
or non-divergent normalization constants.

The Hilbert space extension $\cH=\CC^2\hookrightarrow \CC\PP^2$
together with the observed simultaneous scaling of the whole root
space $\fS_\lb$ obtained in section \ref{jordan}, the upper and
lower triangular (parabolic subgroup type) structure of the
$\fS_\lb-$related matrices in \rf{qa7}, \rf{qa7-b} and the parabolic
subgroup structure \rf{g14} at EPs  provides one more indication
that the natural structure at EPs is connected with projective flags
\cite{parabolic-flag-book}. A study of Jordan chain related flag
bundles and the mappings between their complementary affine charts
will be presented in \cite{G-new1}.

Returning to the $\varepsilon\to 0$ limit in \rf{r3} we see that
\ba{r4}
\frac{c_+^2}{c_-^2}\to -1&\qquad \Longrightarrow \qquad &
\frac{c_+}{c_-}\to \pm i\nn\\
&\qquad \Longrightarrow \qquad & \Phi_+\to \pm i\Phi_-\,,
\ea
i.e. the two eigenvectors (fiber sections) $\Phi_+$, $\Phi_-$ are
phase-shifted one relative to the other by $\pm i$ when tending to
their common coalescence line at $\varepsilon =0$: \ $\pi(\Phi_+)=
\pi(\Phi_-)= \pi(\Phi_0)$. We note that this relative $\pm i$
phase-shift of the vectors $\Phi_+$, $\Phi_-$ is generic for
models in their instantaneous picture and with $d_\pm=c_\pm$ and
normalization $\langle\Xi_\pm|\Phi_\pm\rangle=1.$

A further result which immediately follows from \rf{r3} is the
typical distance-dependent phase jump behavior in the vicinity of
the EP. In a sufficiently close vicinity of an EP
($|\varepsilon|\ll 1$) any sufficiently smooth trajectory in an
underlying parameter space can be roughly approximated by a
straight line segment with an effective parametrization of the
type
\be{a19a}
\varepsilon=e^{i\alpha_0}(\rho + i s),\qquad s\in [-s_0,s_0]\subset
\RR
\ee
where $\alpha_0=$const fixes the direction orthogonal to the
effective trajectory in the complex $\varepsilon-$plane and $\rho$
is the minimal distance $\rho=|\varepsilon (s=0)|$ of this
trajectory to the EP. The parameter along the path is $s\in
[-s_0,s_0]\subset \RR$. This parametrization gives:
\ba{a20}
\left[\varepsilon(s)\right]^{-1/4}&=&
e^{-i\frac{\alpha_0}{4}-i\theta(s)}|\varepsilon(s)|^{-1/4}\nonumber\\
|\varepsilon(s)|&=&\left(\rho^2+ s^2\right)^{1/2} \nonumber\\
\theta(s)&=&\frac 14 \arctan (s/\rho)\in \left(-\pi/8,\pi/8\right)
\ea
and we observe that the minimal distance $\rho$ between the
parameter trajectory and the EP defines the smoothness of the
phase changes. The closer the path approaches the EP the more it
will take the form of a Heaviside step function with jump height
$\pi/4$:
\be{a21}
\theta(s;\rho\to 0)\to \frac \pi 4\left[\Theta (s)-\frac
12\right]\,.
\ee
The phase jump behavior can be used as implicit indicator of a
possible close  location of an EP --- a fact especially useful in
numerical studies of systems with complicated parameter
dependence, but where phases of eigenvectors can be easily
extracted. Jumps $\pm\pi /4$ of wave function phases have been
observed numerically in \cite{ro01} for the model Hamiltonian
(\ref{a1}) and in \cite{rs2} for the special case of a small
quantum billiard. According to these results, the phases of the
components change smoothly (as a function of a certain control
parameter) in approaching the EP and jump by $\pi/4$ at the
smallest distance from this point. Other phase jump values are
possible, but require especially tuned paths.

\section{Phase rigidity\label{rigid}}

In numerical studies of man-made open quantum systems depending in
a complicated way on several parameters $\bX=(X_1,\ldots,X_m)\in
\cM$ it is usually important to know how close a given
configuration is located to an EP. EPs dominate the system
behavior also in their vicinities, spectral bands may merge at EPs
\cite{moiseyev} or the transmission properties of quantum dots
(QDs) may become optimal at EPs \cite{phasrig}. A measure for the
distance between a given point in parameter space and a closely
located EP would provide a convenient tool for adjusting and
tuning parameters so that a system may be 'moved' in parameter
space toward to or away from this EP.

In \cite{phasrig} it has been shown numerically that within the
instantaneous picture ($\Phi=\Xi^*$) an appropriate measure for
the detection of EP vicinities is the fraction
\be{ph1}
r=\frac{\Phi^T\Phi}{\langle\Phi|\Phi\rangle}\,.
\ee
We note that originally similar fractions have been introduced in
\cite{brouwer} to describe the transitions between Hamiltonian
ensembles with orthogonal and unitary symmetry in Hermitian
quantum chaotic systems. There the square modulus $|r|^2$ was
dubbed "phase rigidity". In our considerations of non-Hermitian
systems we use this term in loose analogy for $r$ itself.

Decomposing $\Phi$ into real and complex components
$\Phi=\Phi_r+i\Phi_i$ we find from the normalization that
\be{ph2}
\Phi^T\Phi=1=\Phi_r^T\Phi_r-\Phi_i^T\Phi_i\,,\quad \Phi_r^T\Phi_i=0
\ee
and\footnote{In equation \rf{ph2} it can be set
$\Phi_r^T\Phi_r=:\cosh^2\beta $ and $\Phi_i^T\Phi_i=:\sinh^2\beta$.
This hyperbolic structure shows analogies with the mass shell
condition $E^2-p^2=m^2$ of special relativity. The EP-limit
$\Phi_r^T\Phi_r,\Phi_i^T\Phi_i\to \infty$ corresponds, e.g., to the
light-cone limit where the vectors become isotropic --- a fact which
seems to play an important role in connection with the conjectured
 Hilbert space worm holes \cite{CMB-brachist} related to the
brachistochrone problem of $\cP\cT-$symmetric Quantum Mechanics
(PTSQM).} hence that the norm is bounded below
\be{ph3}
||\Phi||^2=
\langle\Phi|\Phi\rangle=\Phi_r^T\Phi_r+\Phi_i^T\Phi_i=2\Phi_i^T\Phi_i+1\ge
1\,.
\ee
The phase rigidity can be expressed as
\be{ph4}
r=\frac1{||\Phi||^2}\in [0,1]
\ee
where according to \rf{r6} for the EP-limit $\varepsilon\to 0$
holds
\be{ph5}
r\approx |2\varepsilon|^{1/2}\to 0\,.
\ee
The opposite limit $r\to 1$ is reached when the channel coupling
$\omega$ in the Hamiltonian  \rf{a1} vanishes, i.e. when the
interaction between the two decaying resonance states tends to zero
and any eigenvector can be taken purely real-valued in the
instantaneous picture.

Finally, we note that for certain quantum dot systems the phase
rigidity $r$ is closely related to the transmission properties of
these systems. The capability of corresponding numerical studies
(including the visualizations of transmission and phase rigidity
'landscapes' over parameter space) has been recently demonstrated
in \cite{phasrig}.

\section{$\cP\cT-$symmetric models\label{PTSQM}}

Toy model Hamiltonians of  $2\times 2-$matrix type have been often
used as test ground in $\cP\cT-$symmetrically extended Quantum
Mechanics (PTSQM) \cite{CMB-PTSQM,CMB-brachist,CMB-rev}. They can
be obtained from non-Hermitian complex symmetric $2\times
2-$matrix Hamiltonians by imposing a $\cP\cT-$symmetry constraint.
In a suitable parametrization they have the form
\be{pt1}
H=\left(%
\begin{array}{cc}
  re^{i\theta} & s \\
  s & re^{-i\theta} \\
\end{array}%
\right),\qquad r,s,\theta \in \RR
\ee
and commute with the operator $\cP\cT$
\be{pt2}
[\cP\cT,H]=0,\qquad \cP=\left(%
\begin{array}{cc}
  0 & 1 \\
  1 & 0 \\
\end{array}%
\right)\,.
\ee
Here, $\cP$ is the parity reflection operator and $\cT$ --- the
time inversion (acting as complex conjugation). The eigenvalues of
$H$ are
\be{pt3}
E_\pm=r\cos(\theta)\pm \sqrt{s^2-r^2\sin^2(\theta)}
\ee
and the corresponding eigenvectors can be represented as
\cite{CMB-rev}
\ba{pt4}
|E_+\rangle&=&\frac{e^{i\a/2}}{\sqrt{2\cos (\a)}}\left(%
\begin{array}{c}
  1 \\
  e^{-i\a} \\
\end{array}%
\right)=:c_+\chi_+\nn\\ |E_-\rangle &=& \frac{ie^{-i\a/2}}{\sqrt{2\cos (\a)}}\left(%
\begin{array}{c}
  1 \\
 - e^{i\a} \\
\end{array}%
\right)=:c_-\chi_-
\ea
where
\be{pt5}
\sin (\a)=\frac rs\sin(\theta)\,.
\ee
With regard to the indefinite (Krein space type \cite{GSZ-squire})
$\cP\cT$ inner product $(u,v)=\cP\cT u\cdot v$ the vectors are
normalized as
\be{pt6}
(E_\pm,E_\pm)=\pm 1,\qquad (E_\pm,E_\mp)=0\,.
\ee
The indefinite $\cP\cT$ inner product is then mapped by the
dynamical operator $\cC$ with $[\cC,H]=0$ and
\be{pt6-2}
\cC=\frac1{\cos (\a)}\left(%
\begin{array}{cc}
  i\sin(\a) & 1 \\
  1 & -i\sin(\a) \\
\end{array}%
\right)
\ee
(see, e.g., \cite{CMB-rev}) into the positive definite (Hilbert
space type) $\cC\cP\cT$ inner product $((u,v))=\cC\cP\cT u\cdot v$
which yields
\be{pt6-3}
((E_\pm,E_\pm))= 1,\qquad ((E_\pm,E_\mp))=0\,.
\ee

Let us now reshape the model in terms of the EP-relevant notations
of section \ref{setup}. A simple comparison of \rf{a1}, \rf{a4}
with \rf{pt1}, \rf{pt5} shows that
\be{pt7}
Z=i\sin(\a)
\ee
and, hence, that
\be{pt7-2}
\cC=\frac1{\cos (\a)}\left(%
\begin{array}{cc}
  Z & 1 \\
  1 & -Z \\
\end{array}%
\right)
\ee
and that the model is actually one-parametric with essential
parameter $Z$. Together with \rf{a1a} the parametrization \rf{pt7-2}
leads to a representation of the Hamiltonian \rf{pt1} as
\be{pt7-3}
H=E_0I_2+s\cos(\a)\cC\,, \qquad E_0=r\cos(\theta)
\ee
and $[\cC,H]=0$ is fulfilled trivially.

The compatibility of the $\cP\cT$ and the $\cC\cP\cT$ inner
products \rf{pt6}, \rf{pt6-3} with the bi-orthogonality relations
\rf{a3b} is ensured by the fact that for an eigenvector
$\Phi=c(1,b)^T$ exact $\cP\cT$ symmetry requires
$\cP\cT\Phi\propto\Phi$ and, hence, $c^*b^*(1,1/b^*)^T\propto
c(1,b)^T$ so that $|b|^2=1$. For such vectors it holds
$\cP\cT\Phi\propto \Xi^*$ and due to the dynamically tuned $\cC$
also $\cC\cP\cT\Phi\propto \Xi^*$. As result one finds
$\cC\cP\cT\Phi_k\cdot \Phi_l\propto \cP\cT\Phi_k\cdot
\Phi_l\propto \Xi_k^+\Phi_l$ and full compatibility of the
bi-orthogonality with the $\cP\cT$ and $\cC\cP\cT$ inner products
is established.

From \rf{pt7} we see that possible EPs are solely defined by the
value of $\a$. From $Z_c=\pm i$ we find the corresponding critical
$\a_c$ as
\be{pt8}
\a_c=\pm \pi/2+2N\pi,\quad N\in\ZZ\,.
\ee
Furthermore, it follows from \rf{pt5} that a purely Hermitian
model with $\theta=n\pi$, $n\in \ZZ$ corresponds to $\a=N\pi$,
$N\in \ZZ$. Exact $\cP\cT-$symmetry is preserved for $\a\in \RR
-\{\pi/2 +\pi \ZZ\}$, and the corresponding models are
parameterized by elements $Z$ belonging to the purely imaginary
straight line segment connecting the two EPs, i.e. by
$Z\in(-i,i)$, $\Re Z=0$.

According to \rf{pt4}, at the EPs the eigenvectors lay on the same
line $\pi(|E_+\rangle)=\pi(|E_-\rangle)\approx \chi_0=(1,Z_c)^T$ and
their norms diverge for $\a\to \a_c$ like
\be{pt9}
|||E_\pm\rangle||^2=\langle E_\pm|E_\pm\rangle\approx
\frac1{|\cos(\a)|}\to \infty\,.
\ee
The operator $\cC$ in \rf{pt6-2} shows the same singular behavior,
i.e. the $\cC-$induced mapping between the Krein space and the
Hilbert space breaks down at the EPs. In analogy to the
singularity resolution presented in section \ref{instant} we may
map the vectors $|E_\pm\rangle\in \CC^2$ into elements from the
affine chart $U_2\subset \CC\PP^2$ corresponding to points
$e_\pm\in\CC\PP^2$ with homogeneous coordinates
\be{pt10}
|E_\pm\rangle \mapsto e_\pm=\left(\chi_\pm^T,c_\pm^{-1}\right)\,.
\ee
The original normalization via $\cP\cT$ inner product
$\cP\cT|E_\pm\rangle \cdot |E_\pm\rangle=1$ acts then as
generalized conic
\be{pt11}
\cP\cT\chi_\pm\cdot \chi_\pm -\left(\cT
c_\pm^{-1}\right)c_\pm^{-1}=0
\ee
which remains regular in the EP-limit $\a\to\a_c$, but shows the
typical EP-related self-orthogonality (isotropy) of the lines
$\cP\cT\chi_\pm\cdot \chi_\pm\to 0$. Again we arrive at the
conclusion that the original Hilbert space $\cH=\CC^2$ should be
projectively embedded into $\CC\PP^2$ in order to accommodate
EP-related singularities.

Finally, we note that the recently uncovered solutions of the
$\cP\cT-$symmetric brachistochrone problem with vanishing optimal
passage time \cite{CMB-brachist} occurs for $\a=\pi/2 $ what
according to \rf{pt8} can be identified as an EP-regime\footnote{The
corresponding state vector alinement without link to EPs was
observed also in \cite{d-martin}.}. This fact appears compatible
with the results of \cite{fring-brach} where a vanishing passage
time was reported for arbitrary non-Hermitian Hamiltonians. In this
regard it is natural to conjecture that a vanishing optimal passage
time might be a generic EP-related feature of non-Hermitian systems
not necessarily restricted to PTSQM models.

\section{Conclusion\label{conclu}}

In the present paper we considered projective Hilbert space
structures in the vicinity of EPs. Starting from a leading-order
Puiseux-expansion of the bi-orthogonal eigenvectors of a
non-Hermitian (complex symmetric) diagonalizable $2\times 2-$matrix
Hamiltonian in terms of root vectors (algebraic eigenvectors) at an
EP the normalization divergency of the eigenvectors in the EP-limit
has been parameterized. It has been shown that the natural
projective line structure related to the eigenvectors of the
diagonal Hamiltonian has to be replaced at an EP by a higher
dimensional projective structure in which all the root vectors of a
Jordan block scale simultaneously with the same single factor. For a
simplified setup with left eigenvectors equated to their complex
conjugate right counterparts, the normalization divergency has been
resolved by embedding the original Hilbert space $\cH=\CC^2$ into
its projective extension $\cH\hookrightarrow\CC\PP^2$. Eigenvectors
normalized according to the diagonalizable Hamiltonian and
eigenvectors with a normalization inherited from the root vector
normalization live then merely in different (complementary) affine
charts of $\CC\PP^2$. The states themselves can be interpreted as
conics in $\CC\PP^2$. The line structure of the states behaves
smoothly and independently of these charts and their possibly
singular transition functions. This indicates on the possibility of
a technically efficient description of the global behavior of the
non-Hermitian system by factoring the eigenvectors in globally
smoothly varying non-singular projective line components
$\pi(\Phi_k)$ and possibly diverging scale factors\footnote{The
question concerning the physical interpretation of diverging or
non-diverging normalizations and the corresponding diverging or
non-diverging sensitivity in perturbation expansions is highly model
dependent (see e.g. \cite{moiseyev,andrianov-jordan-jpa}) and still
requires a detailed investigation.} $c_k$.

With the help of the Puiseux expanded eigenvectors it has been shown
that the geometric phase obtained on circles around EPs of complex
symmetric Hamiltonians is purely real-valued and that the
corresponding monodromy transformations are induced by an Abelian
parabolic subgroup of $SL(2,\CC)$. Furthermore the Puiseux expansion
has been used to explain phase jumps which in prior work had been
numerically observed in the vicinity of EPs. An analytical
foundation for the usefulness of the phase rigidity as distance
measure to EPs has been provided. Finally, a $\cP\cT-$symmetric
model has been studied. It has been shown that the EP-related
singularities show up not only in the normalization conditions of
the eigenvectors but also in the dynamical symmetry operator $\cC$.
The normalization  singularity has been resolved via a projective
extension of the original Hilbert space. From the singularity
structure it has been conjectured that the zero passage time effect
in the brachistochrone problem of non-Hermitian Hamiltonians might
be a generic EP-related artifact.

{\bf Acknowledgement}.

We thank Hugh Jones and Andreas Fring for useful comments on
\cite{CMB-brachist,fring-brach}. This work has been supported by the
German Research Foundation DFG, grant GE 682/12-3 (U.G.) and by the
grants RFBR-06-02-16719, SS-5103.2006.2 (B.F.S.). \vspace{.5cm}
\appendix
\section{Jordan normal forms for complex symmetric $2\times 2$ matrices\label{jordan-derive}}
At the EPs with $Z_c=\pm i=:\mu i$ the matrix
\be{qa1}
H(Z_c)-E_0 I_2=\omega \left(
\begin{array}{cc}
Z_c &  1 \\ 1 & -Z_c
\end{array}
\right)=:M
\ee
is related to its Jordan normal form $J_2(0)=\left(
\begin{array}{cc}
0 &  1 \\ 0 & 0
\end{array}
\right)$ by a similarity transformation \cite{G-new1}
\ba{qa2}
M=PRJ_2(0)R^{-1}P^{-1}\,.
\ea
From the symmetry properties
\be{qa3}
M=M^T, \qquad J_2(0)=S_2J_2^T(0)S_2
\ee
with $S_2=\left(%
\begin{array}{cc}
  0 & 1 \\
  1 & 0 \\
\end{array}%
\right)$ and $P^2=S_2$ one finds
\ba{qa4}
P= \frac{e^{i\mu\frac\pi 4}}{\sqrt 2}\left(
\begin{array}{cc}
1 &  -i\mu \\ -i\mu & 1
\end{array}
\right),\qquad P=P^T=(P^{-1})^+
\ea
and
\be{qa5}
R= \left(%
\begin{array}{cc}
  q & 0 \\
  0 & q^{-1} \\
\end{array}%
\right),\qquad q:=\sqrt{2\omega}\,.
\ee
The elementary Jordan block $J_2(0)$ has right and left root
vectors $\Theta_0,\Theta_1$ and $\Psi_0,\Psi_1$ satisfying
\ba{qa6}
J_2(0)\Theta_0=0,\qquad J_2(0)\Theta_1=\Theta_0\nn\\
J_2^T(0)\Psi_0=0,\qquad J_2^T(0)\Psi_1=\Psi_0\,.
\ea
The explicit solutions of these Jordan chains can be arranged as
Toeplitz and Hankel matrices
\ba{qa7}
\bfTheta=[\Theta_0,\Theta_1]=\left(%
\begin{array}{cc}
  c_0 & c_1 \\
  0 & c_0 \\
\end{array}%
\right),\quad
\bfPsi=[\Psi_0,\Psi_1]=\left(%
\begin{array}{cc}
  0 & d_0^* \\
  d_0^* & d_1^* \\
\end{array}%
\right)
\ea
and
\be{qa7-b}
\tilde \bfPsi:=\bfPsi S_2=\left(%
\begin{array}{cc}
  d_0^* & 0 \\
  d_1^* & d_0^* \\
\end{array}%
\right)\,.
\ee
From the simplest realization of the bi-orthonormality condition
$\bfPsi^+\bfTheta=S_2$, $\tilde \bfPsi^+\bfTheta=I_2$ one finds
the parameters $c_1=d_1=0$, $d_0c_0=1$. Via similarity
transformations $\Theta_{0,1}\mapsto \Phi_{0,1}=PR\Theta_{0,1}$
and $\Psi_{0,1}\mapsto \Xi_{0,1}=P\left(R^{-1}\right)^+\Psi_{0,1}$
one arrives at the root vectors of $M$
\ba{qa8}
\Phi_0=\sigma qc_0\left(%
\begin{array}{c}
  1 \\
  -Z_c \\
\end{array}%
\right), &\qquad& \Phi_1=\sigma q^{-1}c_0\left(%
\begin{array}{c}
    -Z_c \\
    1 \\
\end{array}%
\right)\nn\\
\Xi_0=\sigma q^*d_0^*\left(%
\begin{array}{c}
    -Z_c \\
    1 \\
\end{array}%
\right), &\qquad& \Xi_1=\sigma q^{*-1}d_0^*\left(%
\begin{array}{c}
        1 \\
        -Z_c \\
\end{array}%
\right)\nn\\
\sigma:=\frac{e^{i\mu\frac \pi 4}}{\sqrt 2}\,.
\ea

\end{document}